\documentclass[twocolumn, nofootinbib, floatfix]{openjournal}
\usepackage{amsmath}
\usepackage{graphicx}
\usepackage{dcolumn}
\usepackage{bm}
\usepackage{epsfig}
\usepackage{amssymb,latexsym,mathrsfs}
\usepackage{graphicx}
\usepackage{color}
\usepackage{hyperref}

\usepackage{float}

\hypersetup{
    colorlinks=true,
    linkcolor=red,
    citecolor=blue,
} 
\usepackage{ulem}
\usepackage{cancel}
\newcommand{\be}{\begin{equation}}
\newcommand{\ee}{\end{equation}}
\newcommand{\bs}{\begin{split}} 
\newcommand{\bea}{\begin{eqnarray}}
\newcommand{\eea}{\end{eqnarray}}

\newcommand{\nflash}{N_{\rm flash}} 
\newcommand{\ntar}{N_{\rm target}} 
\newcommand{\pburst}{p_{\rm burst}} 
\newcommand{\tobs}{t_{\rm obs}} 
 
\newcommand{\tmax}{t_{\rm max}}

\newcommand{\tstar}{t_\star}

\begin{document}

\title{Ultra Fast Astronomy: Optimized Detection of 
Multimessenger Transients} 

\author{Mikhail Denissenya$^1$, Eric V. Linder$^{1,2}$} 
\affiliation{${}^1$Energetic Cosmos Laboratory, Nazarbayev University, 
Nur-Sultan 010000, Kazakhstan\\ 
${}^2$Berkeley Center for Cosmological Physics \& Berkeley Lab, 
University of California, Berkeley, CA 94720, USA
}

%%%%%%%%%%%%%%%%%%%%%%%%%%%%%%%%%%%%%%%%%%%%%%%%%%%%%%%%%%%%%%%%%%%%%%%%
\begin{abstract} 
Ultra Fast Astronomy is a new frontier becoming enabled by improved 
detector technology allowing discovery of optical transients on 
millisecond to nanosecond time scales. These may reveal counterparts 
of energetic processes such as fast radio bursts, gamma ray bursts, 
gravitational wave events, or play a role in the optical search for 
extraterrestrial intelligence (oSETI). We explore some 
example 
science cases 
and their optimization under constrained resources, basically how to  
distribute observations along the spectrum of short duration searches 
of many targets or long searches over fewer targets. 
As a demonstration of the method we  
present some 
analytic and some numerical optimizations, of both raw detections and 
science characterization such as an information matrix analysis of 
constraining a burst delay -- flash duration relation. 
\end{abstract}

\date{\today} 

\maketitle

%%%%%%%%%%%%%%%%%%%%%%%%%%%%%%%
\section{Introduction}

The transient sky is a treasure trove \citep{2001.00588} of events and 
information about the energetic universe, ranging over gamma ray to radio 
wavelengths in light, plus neutrinos and gravitational waves, and over 
time scales from milliseconds to months. Well known cosmological examples 
include fast radio bursts (FRB) and gamma ray bursts (GRB) on millisecond 
to second timescales, binary black hole inspiral gravitational waves over 
milliseconds to days, and supernovae over days to months. Of course the 
entire universe can be viewed as transient over long enough timescales 
\citep{1205.4201,1903.05656,1402.6614,1805.12121,1802.04495,1011.2646}. 

Time domain surveys specifically seek to explore the transient sky, with 
the next generation coming underway \citep{ztf2,lsst}. Continuously scanning 
surveys can also detect transient events 
\citep{2012.14347,1907.04473,1907.12559,1604.03507}. 
Multimessenger detection of 
an event in photons of diverse wavelengths, other high energy particles 
such as neutrinos or cosmic rays, or gravitational waves is a burgeoning 
new field, e.g.\ \citet{1903.09224,2012.02685,1907.07648,1903.08184,1903.08128}. 
However, all these surveys tend to scan over seconds to days, insensitive to 
transients on shorter timescales 
(but see \citet{2101.02454,1905.02429,1008.0605}). 
Certainly we know that astrophysical 
energetic events can occur down to nanosecond timescales, e.g.\ 
\citet{1902.07730,1505.06400}. 
If optical 
and ultraviolet, infrared, etc.\  
surveys sensitive to subsecond 
timescales can be realized, might they reveal a whole new world of events at cosmological distances? 

Detector technology and imaging software is reaching capabilities to 
make such surveys a reality. In particular, the development of silicon  
photomultiplier 
arrays  
(SiPM) with ultrafast readout and coincidence triggers 
\citep{2002.00147,1908.10549} may open up the era of Ultra Fast Astronomy, 
on millisecond and even submicrosecond timescales 
in the optical and near infrared. Other detectors such 
as avalanche photodiodes \citep{1906.03837} are also being explored.  
While they still have a long way to go in terms of large arrays, 
spatial resolution, and noise including crosstalk, they are capable of single photon 
detection and continuous readout. 

The discovery of optical millisecond and below transients 
would be a major breakthrough, rife with astrophysical information 
and numerous applications. Here we focus on one particular aspect 
of this new field of Ultra Fast Astronomy:   
transients visible 
in multiple windows, 
e.g.\ gamma rays and optical light, or gravitational waves and optical 
light, that can be targeted, i.e.\ there is a precursor or 
repeating event. 
That is, we specifically seek connections with known -- or yet 
to be found -- ultra fast transients. Examples include the search for optical 
counterparts of known energetic events, such as repeating fast radio 
bursts (FRB), delayed optical phenomena from a one time event, or even 
as part of an optical search for extraterrestrial intelligence (oSETI) 
program \citep{1808.05772,1603.08928,1907.04443}. 

The size of detectors arrays, the possible field of view, and 
the instrumental characteristics are yet unknown. What we aim for 
here is merely to give a flavor of the science investigations and 
methods that might prove useful. Of necessity these will need to 
be adjusted for specific situations, e.g.\ particular types of bursts 
to follow, instrument properties, etc. However the general 
optimization framework we present should be a good guiding tool, regardless 
of the specific burst properties used as toy examples.  

Framed generally: given a set of adaptable observations, a merit 
function to optimize, a limited resource, and a cost function per 
observation, what is the best strategy for achieving some science 
characterization. Our main topic of investigation will be ways of 
optimizing a detection or characterization of an astrophysical 
process given  
a large set of targets (e.g.\ prior burst locations) 
but limited 
observing resources -- e.g.\ how much time should be  
spent on each target to best achieve the proposed goal. 
We emphasize that this is in essence a follow up program; we assume 
a target list of interesting objects, e.g.\ repeating FRB, exists 
and we seek detection of optical transients from them.  

Astrophysically, such an  
optimization analysis has been carried out for strong gravitational 
lens surveys \citep{linslopt} and supernova surveys (e.g.\ 
\citet{hut2000,0208100}). For example there the 
optimization was for the data set distribution, the 
number of targets followed up at different redshifts, while  
the merit function 
was the dark energy joint parameter estimation uncertainty (``figure of 
merit''), the limited resource was the total 
spectroscopy time, and the cost function expressed the resource use 
of followup spectroscopic time by a target at redshift $z$. 
There may also be a systematics function that 
effectively caps the number of useful 
targets in each observation (e.g.\ redshift) bin. 

To develop and demonstrate the method we explore  
three separate science approaches. 
We emphasize that these are only 
toy examples; applications of Ultra Fast Astronomy are likely to 
be much richer and more varied, and of course instrumental specifics 
will play a much larger role.  
In Section~\ref{sec:frb} we describe the basic 
foundation, in terms of a target source (known for 
repeating or predictable bursts) and its possible optical 
counterparts -- here called flashes. 
Section~\ref{sec:slopen} demonstrates the optimization 
procedure on the question of how to allocate limited 
observing time if one wants to use flash abundances to 
connect flash properties to burst characteristics. 
A more physically incisive test of 
a property such as a burst delay -- flash duration relation, possibly 
useful as a test of o\-SETI, is optimized in 
Section~\ref{sec:taudirect} when using measurements of 
flash durations rather than mere abundances. 
Such examples should give the flavor, and potential exciting 
promise, of the general approach.  
We discuss further potential opportunities from technology 
development and conclude in Section~\ref{sec:concl}. 
In the Appendix~\ref{sec:apxanl} we present some analytic   
results on maximizing detections 
by considering basic two or three time bin optical followup 
programs of repeating bursts, just using this simple example 
to illustrate some general principles of resource constrained 
optimization.

%%%%%%%%%%%%%%%%%%%%%%%%% 
\section{Burst optical counterparts} \label{sec:frb} 

The basic idea is that there is a list of targets for which 
we seek to detect optical transients. In particular, we consider 
the targets to be sources that burst in some other wavelength, 
and we look for optical counterparts. One example might be 
fast radio bursts we follow up with an optical program to look 
for ultra fast optical transients. This phrasing is purely a matter 
of convenience and indication of an interesting science goal, 
without drawing on specific FRB properties --  
the principles are kept generally applicable. 
Fast radio bursts are millisecond transients detected at radio 
wavelengths \citep{1904.07947}. A fraction of them repeat, though not necessarily 
periodically. They may, however, have ``active  windows'' 
that are periodic, with enhanced probability of a repeat burst 
sometime within the window; this has also been detected for 
soft gamma repeaters \citep{dgl}. 

The optical program is thus essentially 
a followup program, scanning targets that have previously burst, 
not a blind survey. We distinguish 
the transients by saying that we look for {\it flashes\/} 
(e.g.\ in optical) that 
are targeted on previous {\it bursts\/} (e.g.\ in radio) -- 
and in particularly targeted at a time when we have some 
reason to expect a repeat burst so that we can look for 
coincident, or nearly coincident, flashes and bursts. (Also, 
active windows for bursts do not always deliver actual activity, 
but it is possible there may be activity of interest at that 
time in other wavelengths.) 

Determining whether FRB, or any other bursting sources, have optical emission, and measuring it, 
would be important for understanding the physics behind the source; 
so far the millisecond 
timescale is too short to make such optical (or UV or NIR)  observations. 
As initial goals of Ultra Fast Astronomy 
we might seek to maximize the chances 
of detecting an optical counterpart, i.e.\ the numbers of such 
burst counterparts, and to learn some basic physics behind the 
flash. We discuss these in 
Section~\ref{sec:slopen} and Section~\ref{sec:taudirect} respectively. 

The first step to investigate the relation between flashes and 
bursts is take a probability for a burst within an observing 
window and propagate it to the number of flashes over a 
distribution of observing times, i.e.\ a survey. 
Suppose 
the probability distribution function (PDF)  
of a burst repeating a time $t$ after a 
previous burst (so we know where to look) and having an 
optical counterpart (so there is something to detect) is 
some normalized $\pburst(t)$. Then in a window of 
time $[t_{i-1},t_i]$ the probability of a viable repeat burst 
is 
\be 
p_i(t_i)=\int_{t_{i-1}}^{t_i} dt\,\pburst(t)\,. \label{eq:probint} 
\ee 
This may be a function of burst properties, and we 
may later be interested in subtypes, but for now we simply 
consider any burst. 

Our basic question is whether we 
want a quick look at many targets or a long look at fewer 
targets -- what is the optimum allocation? 
The observing procedure is taken to be 
assignment from a target list to observing times of various 
durations. 
For example, some targets we will observe for a duration 
$t_1$ and if 
no flash is detected we move on to another target. Some 
targets we will observe for $t_2$ -- if it does not flash 
by time $t_2$ we move on, if it flashes in the first $t_1$ 
time then we do not continue to observe it afterward, and 
instead assign it to the $t_1$ ``bin'', 
and if it flashes 
between times $t_1$ and $t_2$ then it is in the bin 
corresponding to $t_2$, etc. We consider an ordered set of 
discrete times $t_i$, where $t_i>t_{i-1}$ and $t_0=0$. 
This is both for 
mathematical simplicity and because of the possibility that 
instrumental constraints may impose certain intervals. One 
could redo the calculations with finer divisions; this does 
not affect the overall principle. 
The target list should be viewed as locations of interest; 
we do not know exactly when the next event will occur, 
certainly not on the subsecond time scale of the transients. 
Unlike a rolling search for supernova explosions, say, 
where we target 
a set of galaxies, while supernovae will stay visible 
for a month or more, bursts go undetected if we are not already 
observing their location. 

The number of flashes detected is 
\be 
\nflash=\sum_i \ntar(t_i)\,p_i\equiv \sum_i n_i p_i\,,  \label{eq:nfldef} 
\ee 
where $n_i\equiv\ntar(t_i)$ is the number of targets 
with observation duration $t_i$ (i.e.\ producing, or not, a 
flash between $t_{i-1}$ and $t_i$). Again the question we 
focus on is what is the optimum allocation $n_i(t_i)$ -- 
do we want a quick look at many targets or a long look at 
fewer targets? (The following calculations assume a single burst population; at the end of Section~\ref{sec:taudirect} we explore multiple source populations, corresponding to an extra summation index in Eq.~\ref{eq:nfldef} for example.)

A further real world complication is limited resources, 
e.g.\ telescope time, meaning that 
we cannot observe as many targets as 
we want for as long as we want. Each observation comes with a 
cost, taking up part of finite resources $R$. 
The cost function may be simply proportional to  
observing time, i.e.\ how much time we dedicate to searching 
for a flash from a target, so 
\be 
R=\sum_i n_i t_i\,. \label{eq:resdef} 
\ee 
Here we take all times, and $R$, to be in units of $t_1$, 
the shortest observation time. That may be  
one second, one day, or whatever; that will depend on the 
specific target population and instrument, but we 
leave that for future detailed survey design. 
Given detailed instrument and survey design, the observing procedure can be further adjusted by employing Monte Carlo methods taking into account, e.g., telescope slew and readout times (see the discussion on such issues at the end of Section~\ref{sec:taudirect}). We also emphasize that at this stage we are not aiming for a complete survey, detecting all counterparts, but rather a reasonable attempt at an initial set of counterpart flashes.

%%%%%%%%%%%%%%%%%%%%%%%%%%%%%%%%%%%%%%%% 
\section{Delay-Duration Relation -- via Abundance} \label{sec:slopen} 

While detecting ultra fast optical transient counterparts 
would be exciting and significant, it would be even better to 
extract some physics out of the detections. 
We consider the following as one illustration. 
Suppose the optical 
flash durations $\tau$ are connected with the burst repeat delay time $t_p$ through a power law scaling. 
We write  
\be 
\tau=A\tstar\,\left(\frac{t_p}{\tstar}\right)^s\,, 
\label{eq:tautp} 
\ee 
where $\tstar$ is a fixed pivot scale, $A$ is the dimensionless 
amplitude of the relation (with a $\tstar$ out front for correct 
dimensionality) and $s$ is the slope relating the mean burst 
delays to the flash durations. 
This could be relevant to the astrophysical bursting mechanism,  
e.g.\ of FRB or GRB, and also to oSETI, where a 
civilization might choose to broadcast short messages 
frequently, or long messages infrequently, in a sort of 
constant energy output ($s=1$) schema. 
We would like to test such an assumption, e.g.\ 
is $s\ne0$ so there is a relation between the burst 
delay $t_p$ and the flash duration  $\tau$, and characterize it -- what estimates of 
amplitude $A$ and slope $s$ do we obtain from data.  

We will use an information matrix approach to 
determining the physical parameters $A$ and $s$ from the 
flash detections, and then optimizing the 
observation time distribution to get the best constraints. 
In this section we take the observable to be the number of 
flashes in a certain observation time bin $t_i$, 
summed over the flash durations $\tau$. 
We want to relate these 
measured abundances to the underlying process parameters 
of the delay-duration amplitude $A$ and slope $s$. 
(In Section~\ref{sec:taudirect} we take a more direct approach, using the flash durations themselves.) 

We begin with the number of flashes with duration $\tau$ detected from all bursts observed with  
observation 
time duration in bin $i$: $t_{i-1}<t_{\rm obs}<t_i$. 
A subscript $i$ means such a time bin and we take all bins to be of equal, unit width. Then 
\bea 
N_{\rm flash}(t_i;\tau)&=&n_{\rm burst}(t_i)\, p_{\rm flash}(t_i;\tau)\\ 
&=&n_{\rm burst}(t_i)\,\,p_{\rm  flash}(t_i;t_p)\,\left|\frac{dt_p}{d\tau}\right|\,, 
\eea  
where we used $p_{\rm flash}(t_i;\tau)d\tau=p_{\rm flash}(t_i;t_p)dt_p$. The quantity $p_{\rm flash}(t_i;t_p)$ is taken to be the probability of 
the repeat burst within the observing time, i.e.\ 
Eq.~\eqref{eq:probint}, times some probability for 
having a flash associated with a burst, $p_\star$. 
We set $p_\star=1$ for simplicity; a constant $p_\star$ 
does not affect the form of our results. 

We consider a Gaussian delay model where 
\be 
\pburst(t)=\frac{1}{\sqrt{2\pi\sigma^2}}\,e^{-(t-t_p)^2/(2\sigma^2)}\,.  
\label{eq:pburst}
\ee 
This represents a burst that occurs a mean time $t_p$ after 
time zero (e.g.\ the previous burst or the beginning of an 
activity window), but with some 
uncertainty $\sigma$. 
While this is just a model, it is a reasonable first 
choice; one would 
need to determine the burst delay distribution in the process of building up  
the target list. 
The probability for a flash within a window $[t_{i-1},t_i]$, and hence being assigned to observation duration $t_i$,   
is then  
\be 
p_i(t_i)=\frac{1}{2}\,\left[{\rm erf}\left(\frac{t_i-t_p}{\sigma\sqrt{2}}\right)-{\rm erf}\left(\frac{t_{i-1}-t_p}{\sigma\sqrt{2}}\right)\right]\,. 
\label{eq:piti}
\ee 

This gives all the necessary ingredients for the 
calculation. 
Using the inverse of Eq.~\eqref{eq:tautp}, 
\be 
t_p=\tstar\,\left(\frac{\tau}{A\tstar}\right)^{1/s}\,, 
\label{eq:tptau} 
\ee 
we can calculate the Jacobian   
\be 
J\equiv \left|\frac{dt_p}{d\tau}\right|=\frac{1}{As}\,\left(\frac{\tau}{A\tstar}\right)^{(1-s)/s}\,. \label{eq:jac} 
\ee 
Explicitly,  
\bea  
N_{\rm flash}(t_i;\tau)&=&n_{\rm burst}(t_i)\,\,p_{\rm  flash}(t_i;t_p(\tau,A,s))\notag\\ 
&\qquad&\times\frac{1}{As}\left(\frac{\tau}{A\tstar}\right)^{(1-s)/s}\,. 
\eea 

To carry out the information analysis, we need the 
sensitivities: the derivatives with respect to the parameters,  
\be 
\frac{\partial\nflash(t_i,\tau)}{\partial A}\,,\qquad 
\frac{\partial\nflash(t_i,\tau)}{\partial s}\,.  
\ee 
Note that 
\be  
\frac{\partial p_i(t_p(\tau,A,s))}{\partial\theta}=\frac{\partial p_i(t_p(\tau,A,s))}{\partial t_p}\,\frac{\partial t_p}{\partial\theta}\,,  
\ee 
for $\theta=\{A,s\}$ and 
$p_i(t_p)\equiv p_{\rm  flash}(t_i;t_p)$. 
The derivatives  
\be 
\frac{\partial t_p}{\partial A}=\frac{-t_p}{sA}\,,\qquad 
\frac{\partial t_p}{\partial s}=\frac{-t_p}{s^2}\,\ln\left(\frac{\tau}{A\tstar}\right)\,, 
\ee 
and for the Jacobian factor 
\be    
\frac{\partial J}{\partial A}=\frac{-J}{sA}\,,\qquad 
\frac{\partial J}{\partial s}=\frac{-J}{s^2}\,\left[s+\ln\left(\frac{\tau}{A\tstar}\right)\right]\,. 
\ee   
Putting it all together, 
\be 
\frac{\partial\nflash(t_i,\tau)}{\partial A}= 
\frac{-\nflash}{sA} \left[\frac{\tstar}{p_i(t_p)}\frac{dp_i}{dt_p} \left(\frac{\tau}{A\tstar}\right)^{1/s}+1\right]\,, 
\ee 
and 
\bea 
\frac{\partial\nflash(t_i,\tau)}{\partial s}&=& 
\frac{-\nflash}{s^2} \left[\frac{\tstar}{p_i(t_p)}\frac{dp_i}{dt_p} \left(\frac{\tau}{A\tstar}\right)^{1/s}\ln \left(\frac{\tau}{A\tstar}\right)\right.\notag\\ 
&\quad&\left.+s+\ln \left(\frac{\tau}{A\tstar}\right)\right]\,, 
\eea 
where the derivatives are evaluated at $t_{p,{\rm fid}}=t_\star(\tau/A\tstar)^{1/s}$. For $p_i(t_p)$ given by 
Eq.~\eqref{eq:piti}, its derivative is a difference between Gaussians; note the $1/p_i(t_p)$ term cancels the $p_i(t_p)$ in 
$N_{\rm flash}$. 

We also need to specify the abundance measurement noise matrix, e.g.\ 
Poisson measurement error on the abundances so that $C^{-1}=\nflash(t_i,\tau)$, and the fiducial values, e.g.\ 
$s=1$ and $A=1/4$, with a pivot scale $\tstar=40$. The  
information matrix is summed 
over the bins in $\tau$ 
-- with the constraint that $\tau<t_i$ 
(we can't measure a duration longer than we have observed for) 
-- and then summed over bins in observing time $t_i$. 
Thus, 
\bea 
F_{pq}&=&\sum_{{\rm bins\  of\ }t_i}\ \sum_{{\rm bins\ with\ }\tau<t_i}\notag\\ 
&&\quad\frac{\partial \nflash(t_i,\tau)}{\partial \theta_p}\frac{\partial \nflash(t_i,\tau)}{\partial \theta_q}\,\nflash(t_i,\tau)\,. 
\label{eq:fisher3}
\eea 

\begin{figure} 
\centering 
\includegraphics[width=\columnwidth]{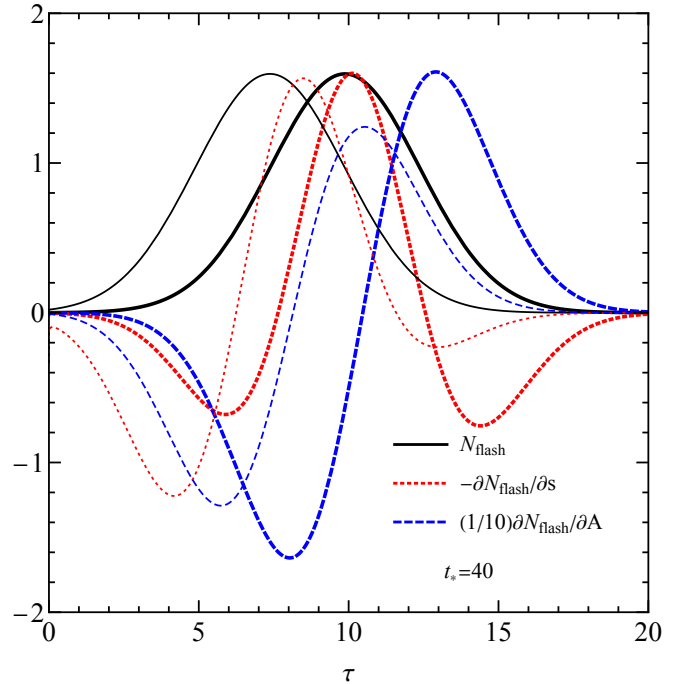}
\caption{The sensitivity derivatives $\partial \nflash/\partial\theta$ (dotted red for $s$, dashed blue for $A$) have 
very different shapes, i.e.\ dependence on the 
observed flash duration $\tau$, so we may expect 
little covariance between them, allowing both  
to be well estimated from data. The solid black curves 
show $\nflash(t_i,\tau)$; the dashed blue (dotted red) curves show the derivative with respect to the amplitude $A$ (slope $s$). Bolder curves are for $t_i=40$, 
thinner curves are for $t_i=30$. 
} 
\label{fig:sens} 
\end{figure}

Figure~\ref{fig:sens} 
shows $\nflash(t_i,\tau)$ and the  
sensitivity derivatives $\partial\ln\nflash/\partial\theta$ for 
our parameters, $\theta=A$ and $s$. 
We see the sensitivity curves vs the flash duration 
$\tau$ have very different shapes, implying little 
covariance is expected in their determination. Both the 
delay-duration amplitude $A$ and slope $s$ should be 
well determined if we observe a range of durations.

%%%%%%%%%%%%%%%%%%% 

\begin{figure} 
\centering 
\includegraphics[width=\columnwidth]{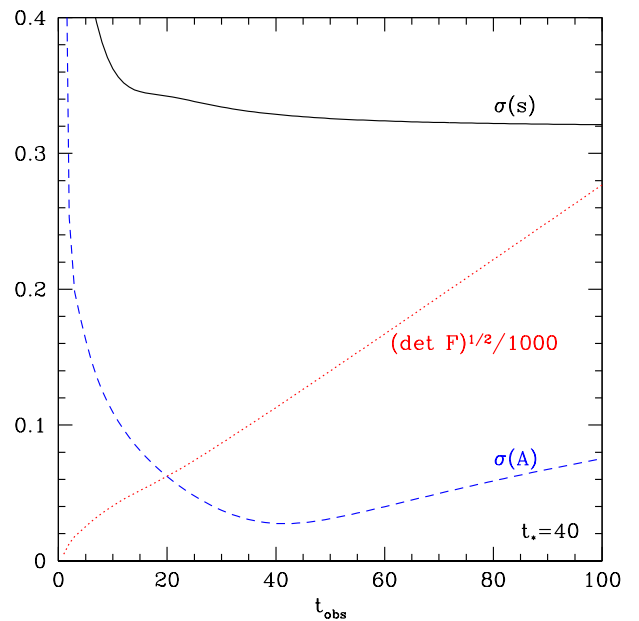}
\caption{Parameter estimation and figure of merit as a function of observing time bin $t_i\equiv \tobs$. 
For all bins $n_i=10$ here. 
} 
\label{fig:tisigmas} 
\end{figure}

Figure~\ref{fig:tisigmas} illustrates the information 
analysis results for the parameter uncertainties and the 
figure of merit $(\det F)^{1/2}$  related to the inverse 
area of the amplitude--slope confidence contour. We see that 
the slope is determined at about the same level for 
$\tobs\gtrsim10$ while the amplitude is best determined near 
the pivot scale $\tstar=40$. Due to the changing covariance 
between the parameters, however, the figure of merit (FOM) 
ends up being monotonic, gaining more information with a longer 
observation time.

%%%%%%%%%%%%%%%%%%%%%%%

\begin{figure} 
\centering 
\includegraphics[width=\columnwidth]{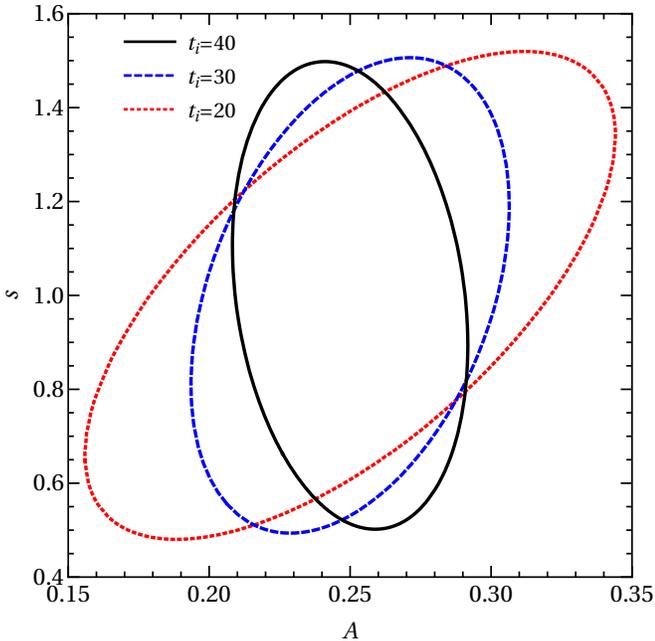}
\caption{Joint probability contours (68.3\% confidence level) for the amplitude $A$ and slope $s$, for three different values of observing time bin $t_i$. 
For all bins $n_i=10$ here. 
} 
\label{fig:ellipnonopt} 
\end{figure}

Figure~\ref{fig:ellipnonopt} shows the joint $1\sigma$ 
confidence contours for three different values of $\tobs$. 
As expected from Fig.~\ref{fig:tisigmas}, the estimation of the slope 
$s$ changes little, but the probability ellipse rotates to 
give a smaller range for amplitude $A$ near $t_i=40$, thinning the ellipse as $t_i$ 
increases, giving a smaller area and larger FOM.

%%%%%%%%%%%%%%%%%%%%%%%

For the full information analysis, one sums over 
the whole range of observing time bins $t_i$, weighted by 
the number of targets in that bin, $n_i$. This will of 
course reduce the uncertainties on the parameter 
estimation. However our main aim is to determine the optimum 
observing strategy, i.e.\ the optimum distribution $n_i$, 
under our observing time resource constraint and cost model. 

We use two independent codes to carry out the optimization. 
One is the information analysis optimization code described in 
\citet{linslopt}, adapted for the present observables and 
variables, that evaluates the change in merit with resource 
amount (observing time), bin by bin, selects the bin with weakest 
leverage, reduces that $n_i$, and reallocates its time to other 
bins (hence changing $n_j$) in a fashion weighted by the merit 
per resource used. 
The other uses the interior point optimization algorithm which combines the resource constraint and the merit function using the barrier function approach \citep{intpmeth}. They provide valuable crosschecks on results and convergence, and we find excellent agreement 
between them. 

The results show this optimization favors maximum numbers in 
the lowest observation time bin, due to the resource 
constraint. 
While FOM improves with $\tobs$, lower $\tobs$ 
allows many more targets within the given resource constraint.  
Since the information matrix basically goes as $n_i^3$ 
(two factors from the two observable $\nflash$ proportional to 
$n_i$ and one from the Poisson noise model), putting all 
observations in the lowest $\tobs$ bin wins. Of course there 
may be other constraints limiting the number of targets, such 
as from available repeaters, the telescope,  detectors, etc. 
In that case, the 
optimum found fills up the lowest bin to the maximum allowed, 
proceeds to the next lowest bin, and so forth until the 
resource allotment runs out. 

A short analytic proof of the behavior comes from considering 
the FOM if all observing is in bin $p$ vs $q>p$. In the first 
case, ${\rm FOM}_1=cn_p^3 p$, where from Fig.~\ref{fig:tisigmas} we 
approximate the dependence of FOM on $\tobs$ as roughly linear. 
In the second case ${\rm FOM}_2=cn_q^3 q$, but the resource constraint 
imposes that $n_q=n_p p/q$. Therefore we have 
${\rm FOM}_2={\rm FOM}_1\, (p/q)^2<{\rm FOM}_1$ since $p<q$. Thus the lower bin always 
wins for this model with the abundance as the main observable.

%%%%%%%%%%%%%%%%%%%%%%%%%%%%%%%%%%%%%%%%
\section{Delay-Duration Relation -- via Direct Measurement} \label{sec:taudirect} 

The optimization in the previous section led to an all or 
nothing result: the solution involved flitting from target to 
target as fast as possible to observe as many as one could 
within the resource constraint. This was driven by the strong 
dependence of the information matrix on the number of events, 
due to taking the abundance as the central observable. A more 
subtle, and interesting, result comes if the observable is 
the flash duration itself, for deriving the relation to the 
burst delay. We analyze this case here. 

The information matrix now becomes 
\bea 
F_{jk}&=&\sum_{t_i}\ \sum_{\tau<t_i} 
\frac{\partial\tau}{\partial \theta_j}\frac{\partial \tau}{\partial \theta_k}\,\nflash(t_i,\tau)\,C^{-1}\\ 
&=&\sum_{t_i}n_i\ \sum_{\tau<t_i} 
\frac{\partial\tau}{\partial \theta_j}\frac{\partial \tau}{\partial \theta_k}\,p(t_i,t_p(\tau))\left|\frac{dt_p}{d\tau}\right|\,C^{-1}
\,, 
\label{eq:fishertau}
\eea 
where $C$ is the duration measurement noise matrix. The sensitivities 
are 
\be 
\frac{\partial\tau}{\partial A}=\frac{\tau}{A}\,,\qquad 
\frac{\partial\tau}{\partial s}=\frac{\tau}{s}\,\ln\left(\frac{\tau}{A\tstar}\right)\,, 
\ee 
the Jacobian $|dt_p/d\tau|$ is given by Eq.~\eqref{eq:jac}, 
and for $C$ we take a diagonal matrix with elements 
$C=\sigma^2_{\tau}\,\delta_{qq}$ where 
$q$ is the index for the $\tau$ bin and 
\be 
\sigma^2_\tau=\sigma^2_{\rm stat}+\nflash(t_i,\tau)\,\sigma^2_{\rm sys}\,. 
\ee 
The statistical contribution may depend on $\tau$, though 
here we take a fiducial case $\sigma_{\rm stat}=1$; the 
systematic term imposes a floor on the measurement uncertainty, or 
ceiling on the number of events such that increasing their quantity 
does not significantly add to  
the information beyond some limit (i.e.\ it cancels 
the $\nflash(t_i,\tau)$ in the numerator of $F_{jk}$: effectively, 
instead of a measurement uncertainty $\sigma_{\rm stat}/\sqrt{\nflash}$ one has 
$\sigma_{\rm  sys}$). We investigate two 
fiducials, $\sigma_{\rm sys}=0$ or 1, as the simplest 
possible examples. 
Real systematics would be determined by the experiment 
design, involving instrumentation properties related to 
readout, crosstalk, etc. Ultra Fast Astronomy instrumentation 
is not yet sufficiently advanced to yield an   
estimate of such systematics so we use our naive model to 
illustrate how it enters the optimization method we present.  

\begin{figure} 
\centering 
\includegraphics[width=\columnwidth]{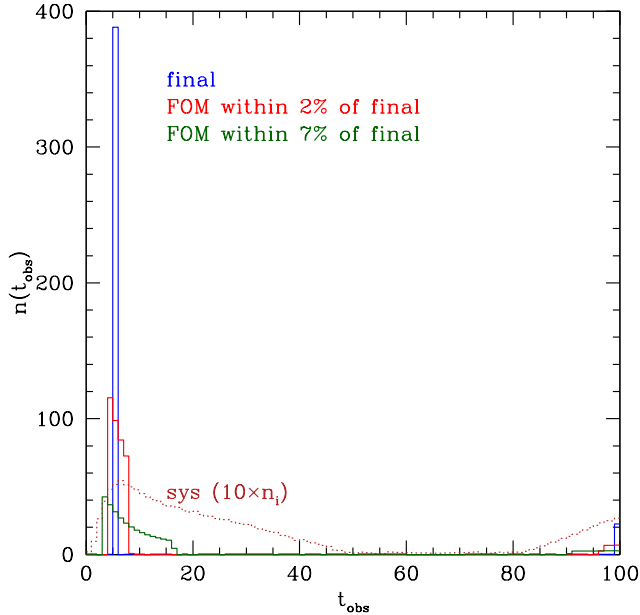}
\caption{Histogram of the optimized target distribution $n_i$, 
producing the maximum FOM under a fixed resource constraint. The solid histograms are for a purely 
statistical measurement uncertainty of $\sigma_{\rm stat}=1$. Distribution results $n_i$ are shown for three  
different stages in the optimization process approaching the final convergence having sharp peaks at $t_i=6$ and $t_i=100$. Parameter constraints  
generally have smaller deviations from their final values than that listed for FOM. The dotted 
brown curve shows the impact of turning on a systematic, 
with $\sigma_{\rm sys}=1$, where the optimized histogram is plotted at ten times its true height (i.e.\ really $n_i<6$). 
} 
\label{fig:histotau} 
\end{figure}

We start the optimization from  an initial state of uniform 
$n_i$ across all bins $t_i=[1,100]$, giving a fixed resource constraint of $R=5050$. 
Of course the survey becomes less sensitive to sources with delay times near or beyond the limit, but we want to maintain a broad survey. 
First we take a 
purely statistical uncertainty, with $\sigma_{\rm stat}=1$. 
Parameter estimation results will scale linearly with this, and as the 
reciprocal square root of the total resources available; 
however it will not affect the form  of the distribution optimization for $n_i$ 
(for $\sigma_{\rm stat}$ independent of $\tau$). 
The final converged optimization gives two peaks in the $n_i$ target 
distribution, at $t_i\approx6$ and 100, as seen in 
Fig.~\ref{fig:histotau}. 
Note \citet{hut2000} also found that for 
a system with $P$ parameters the distribution is optimized with 
$P$ delta functions. The numerical solution approaches isolated 
peaks, with earlier iterations 
having slightly broader peaks, 
but these possess very close to the final parameter uncertainties and FOM  
so we see that small variations in the distribution (e.g.\ due to observational requirements) would have only a minor impact on the science results.

%%%%%%%%%%%%%%%%%

A systematic measurement uncertainty on $\tau$, given by 
$\sigma_{\rm sys}$, will effectively place a ceiling on the number of 
productive targets in a given observing bin $t_i$. This 
flattens the distribution peaks, broadening them and 
distributing the targets over many more bins. Including 
the systematics in quadrature, with $\sigma_{\rm sys}=1$, 
worsens the parameter estimation by 22\% on $A$ and 44\% on $s$, 
and by 33\% on FOM.

\begin{figure} 
\centering 
\includegraphics[width=\columnwidth]{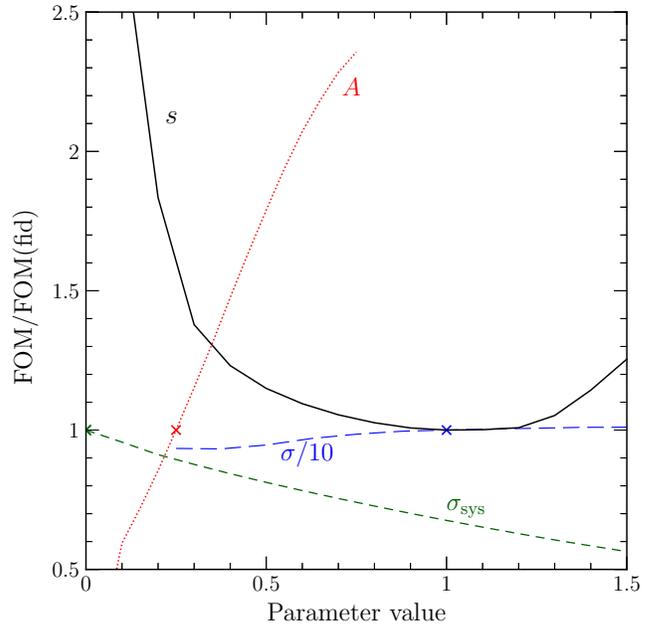}
\caption{The change of FOM relative to the fiducial result is 
plotted vs the parameter values away from their fiducials 
(marked with x's): 
$s_{\rm fid}=1$, $A_{\rm fid}=0.25$, $\sigma=10$, $\sigma_{\rm sys}=0$. 
To keep $\sigma$ on the same scale the x-axis is $\sigma/10$; 
note we cut off the curve for small $\sigma$ as it approaches 
the time bin width. For $A$ we cut off the curve as the flash 
duration approaches the burst delay time  $A=1$, otherwise it 
would not be a flash but a continuous glow. 
} 
\label{fig:fompar} 
\end{figure}

We study the sensitivity of the FOM to the fiducial parameters 
in Fig.~\ref{fig:fompar}. Our fiducial value for the slope, $s=1$, 
is seen to be the most conservative choice: other values of $s$ 
give a higher FOM and tighter parameter estimation. This makes 
sense in that, broadly, when $s=1$ some parameter dependence does 
not enter, e.g.\ in Eq.~\eqref{eq:jac} and hence $\nflash$. More 
specifically, for smaller $s$ we also amplify the sensitivity 
$\partial\tau/\partial s$ and increase the Jacobian $dt_p/d\tau$. 
Thus we expect small $s$ to show the greatest improvement in 
parameter constraints and FOM. What we find for $s>1$ is that, while the 
parameter constraints weaken, the FOM increases since there is 
less covariance between parameters -- this is due to the enhanced range of 
$\tau$ contributing to a given favored $t_p$ near $t_i$ from the 
$1/s$ exponent in the relation Eq.~\eqref{eq:tptau}.

%%%%%%%%%%%%%%%%%

For the amplitude $A$, the main effect is that larger $\tau$ is 
preferred for a given $t_p$ near $t_i$. Since the sensitivities 
strengthen with increased $\tau$, and the information goes 
as the product of sensitivity factors, this overcomes the reduced 
$dt_p/d\tau$. In fact, since $\partial\tau/\partial s$ is the only 
sensitivity benefiting (with the increase in $A$ and $\tau$ 
canceling in $\partial\tau/\partial A$), the improvement in the 
estimation of $s$ and in FOM is roughly linear in $A$, while the 
estimation of $A$ weakens slightly. 
(Only FOM is shown in Fig.~\ref{fig:fompar}.) 

Increasing the systematic $\sigma_{\rm sys}$ of course decreases 
FOM and worsens the parameter estimation. Even within 
the naive model one can get a sense, at least qualitatively, 
for how systematics 
can impact the results. The uncertainty in the 
burst wait time, $\sigma$, has a minor effect, since it only enters 
into the probability factor and not the sensitivities. 
Decreasing $\sigma$ makes it a little harder for $t_i$ to match 
$t_p$, but we can still find some $t_i$ matching $t_p$, where the 
probability is maximal. 
We have verified that varying $\sigma$ continues to produce a sharp peak at $t_i=7$ or slightly earlier while preserving another peak at $t_i=100$.

\begin{figure} 
\centering 
\includegraphics[width=\columnwidth]{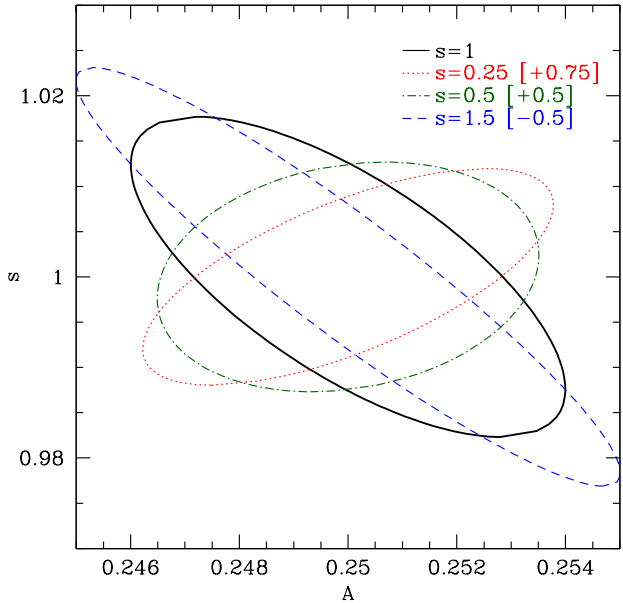}
\caption{$1\sigma$ joint confidence contours in the $A$--$s$ 
plane, for different fiducial values of $s$. Constraints are for 
statistical uncertainties only with $\sigma_{\rm stat}=1$. 
Contours have been shifted vertically by the amount in brackets 
so that all are centered at $s=1$, for easier comparison, despite 
the different fiducial values for $s$.  
} 
\label{fig:ellas} 
\end{figure}

Figure~\ref{fig:ellas} shows how the parameter constraints evolve 
as we change the fiducial slope $s$, as well as the impact on the 
FOM (inverse area of the confidence contour). In particular, 
we can see how for $s=1.5$, although the individual parameter 
constraints are weaker, the FOM is higher due to the narrowing 
of the joint contour. As the fiducial $s$ changes, the 
covariance between $A$ and $s$ does as well, rotating the 
contour and squeezing or broadening it. The slope and amplitude 
of the burst delay -- flash duration relation can be determined 
at the percent level, within this model, for 
the resources assumed. 
This gives the flavor of how such a result could be  
promising for testing, e.g., an oSETI hypothesis of 
constant energy output, i.e.\ $\tau/t_p={\rm constant}$, or $s=1$. 
Since this plot is with statistical 
uncertainties only, the constraints scale with resources: 
linearly for FOM and as the inverse square root for the parameter 
estimation uncertainties.

%%%%%%%%%%%%%%%%%

Finally, suppose we have multiple source populations with 
different properties. While the sum over flash duration $\tau$ 
in Eq.~\eqref{eq:fishertau} does sum over a range of burst durations by 
Eq.~\eqref{eq:tptau}, different populations may have different 
relations. We therefore now consider two populations, with 
relations having characteristics $\tstar=40$ or $\tstar=20$, 
and determine how the optimization changes.  

\begin{figure} 
\centering 
\includegraphics[width=\columnwidth]{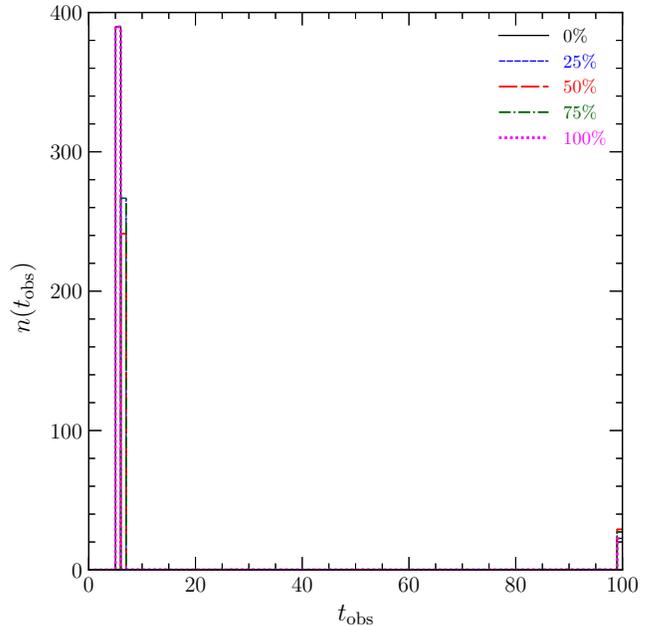}
\caption{Histogram of the optimized target distribution $n_i$, 
producing the maximum FOM under a fixed resource constraint, 
for the case of two source populations. Different fractions 
of the source population of $\tstar=20$ targets relative 
to the full two source population of $t_\star=20$ and 
$\tstar=40$ targets have nearly the same optimized 
distribution. 
Here $\sigma_{\rm stat}=1$, $\sigma_{\rm sys}=0$.  
} 
\label{fig:histts} 
\end{figure}

Figure~\ref{fig:histts} varies the $\tstar=20$ population 
from 0\% of the total (as used in the previous calculations) to 
100\%. We see this has very little impact on the observation 
duration optimization: the preferred survey still has a two peak 
distribution of times, one at the maximum duration and one at 
much shorter (but not minimum) duration. The lower peak does 
shift a little from $\tobs=6$ to 7, depending on the population 
fraction, but we saw in Fig.~\ref{fig:histotau} that the 
figure of merit is fairly forgiving of slight deviations from 
the formal optimum.

%%%%%%%%%%%%%%%%%%%

\begin{figure} 
\centering 
\includegraphics[width=\columnwidth]{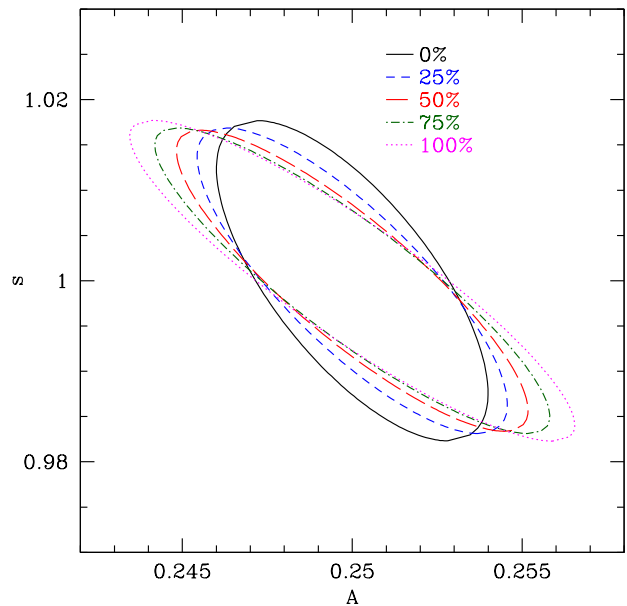}
\caption{$1\sigma$ joint confidence contours in the $A$--$s$ 
plane for the two source population case, with different 
fractions of the source population of 
$t_\star=20$ targets relative 
to the total sum of $t_\star=20$ and $t_\star=40$ targets. 
Constraints are for 
statistical uncertainties only with $\sigma_{\rm stat}=1$.  
} 
\label{fig:elltsas} 
\end{figure}

Figure~\ref{fig:elltsas} shows that the constraints on 
the parameters of the flash duration -- burst delay relation 
do change somewhat with source population fraction. The 
amplitude uncertainty $\sigma(A)$ shows a 62\% variation from 
fraction 0 to 1, while the slope uncertainty $\sigma(s)$ only 
varies by 6\%; the covariance between the two is important 
and compensates such that the FOM only varies by 16\%.

%%%%%%%%%%%%%%%%%

A technical question we have not addressed concerns scheduling. 
How should one decide which targets to observe at what time, 
especially as observing one target may mean missing a favored 
time for another target? To the extent that all targets are 
equal, for example during the first exploratory survey when 
it is not known if any have optical counterparts to follow up, 
then a random choice as used here is fine, or one might focus on targets 
at the best airmass for better signal to noise, or close 
together for reduced telescope slew time. When UFA detectors 
advance to the state when large arrays can cover a wide field, 
then several targets can be followed simultaneously. 
As mentioned in Sec.~\ref{sec:frb}, we have not aimed for a complete survey but one with a promising and interesting set of initial detections. 
Other 
considerations may be a desire for a redshift distribution or 
concentration on, say, galaxy clusters. These are all important 
questions and as we learn more about the source populations 
and instrument characteristics the survey planning would extend 
to these issues.

%%%%%%%%%%%%%%%%%%%%%%%%%%%%%%% 
\section{Conclusions} \label{sec:concl} 

Ultra Fast Astronomy has the potential to uncover a new view of the universe, 
unknown to surveys scanning celestial sources on time scales of  
seconds or longer. Technology such as arrays of photomultipliers and 
photodiodes is approaching the capability of millisecond, 
microsecond, or even nanosecond resolution in the optical 
and possibly NIR time domain. 

This may reveal not only new classes of transients, but 
connections between transients across wavelengths and 
multimessenger signals. Such detections could provide critical 
clues to the physical origins and mechanisms behind highly 
energetic events -- or even hints in the search for 
extraterrestrial intelligence. 

We present a method and explore three cases of optimizing observations so as to  
maximize science under constrained resources, such as telescope 
time. 
We investigate science characterization from counterpart 
(flash) detection, such as testing a burst delay -- flash 
duration relation. First we analyze optimizations in terms of 
measured abundances, and then using measured flash durations. 
The study includes interpretation of the parameter estimation 
constraints and figure of merit (inverse area of the joint 
confidence contour), scaling with intrinsic and survey parameter 
values, and role of statistical and systematic uncertainties. 
In the Appendix, 
we look at optimizing the number of counterpart 
detections, and solve analytically a toy model of how to 
distribute search durations before moving to another target, just using this simple example 
to illustrate some general principles of resource constrained 
optimization. 

Under constrained resources, the derived optimum 
for the delay-duration model studied in the most physically 
incisive case, Sec.~\ref{sec:taudirect},   
is a two prong 
observational strategy of many targets observed for a short 
(but not minimal) time and a few targets observed for a 
(maximally) long time. 
This continues to hold when including multiple source 
populations, at least for the simple two population case we 
calculated.  
Resulting parameter estimation can 
reach the percent level. For example, a ``constant energy 
output'' schema of extraterrestrial signaling could be 
measured at signal to noise $S/N\approx100$ for our fiducial 
resource level. 

These are examples of survey strategy and science analysis 
that could be useful for Ultra Fast Astronomy. 
They are meant to be illustrative, and hopefully 
inspirational, of methods and potential 
science goals, not (yet) rigorous or comprehensive. 
Throughout, we assume an established target list of burst 
sources (e.g.\ in the radio or X-ray) that we want to scan 
for optical flashes. The specific optimization depends on 
having guidance on the burst repeat time probability distribution 
function, e.g.\ whether it favors rapid repeats, long term 
ones, or ones tending to occur in some characteristic  window. 
Thus the Ultra Fast Astronomy survey would be matched to the 
burst populations being followed up. One could alternately 
of course carry out a blind scan of the sky or target interesting 
nontransient or nonrepeating sources; all these types of surveys 
will no doubt play a role in Ultra Fast Astronomy. 
Another important aspect of future work that will go hand 
in hand with the technology development is connecting the 
measurement uncertainty model more 
closely with promising detector technology characteristics. 
This is a new frontier, and there are many directions to explore.

%%%%%%%%%%%%%%%%% 
\acknowledgments 

We thank Albert Wai Kit Lau for helpful discussions. 
This work is supported in part by the Energetic Cosmos 
Laboratory. 
EL is supported in part by the 
U.S.\ Department of Energy, Office of Science, Office of High Energy 
Physics, under contract no.~DE-AC02-05CH11231.

%%%%%%%%%%%%%%%%%%% 
\appendix 

\section{Analytic Cases} \label{sec:apxanl} 

Here we consider a much more restricted question than in the 
main text, but one useful as an analytic example. If our main 
focus is purely detecting a maximum number of flashes, without 
dealing with their properties, then we seek to maximize $\nflash$. 
Combining Eqs.~\eqref{eq:nfldef} and \eqref{eq:resdef}, we have 
\be 
\nflash=p_1\left[\frac{R}{t_1}-\sum_{i>1} n_i\left(\frac{t_i}{t_1}-\frac{p_i}{p_1}\right)\right]\,. 
\label{eq:nflashsum} 
\ee 

To maximize $\nflash$ with respect to the distribution $n_i$, we evaluate the 
partial derivative, giving a critical condition 
\be 
\frac{t_i}{t_1}=\frac{p_i}{p_1}\,. \label{eq:tpequal}
\ee 
This is readily understandable as the boundary 
determining the sign of the 
term with the minus sign prefactor in Eq.~\eqref{eq:nflashsum}, 
thereby increasing or decreasing $\nflash$. 

There are three cases. One is that this has no solution 
because while by construction $t_i>t_1$, we may have 
$p_i<p_1$, depending on the form of $\pburst(t)$. In this 
case the quantity in parentheses in Eq.~\eqref{eq:nflashsum} 
is positive and so the maximum $\nflash$ is when the  
sum goes to zero. That is, the optimum (in terms of many 
detections, not the science property characterization of 
the main text) is to only observe 
on the shortest timescale, $t_1$, with $n_{i>1}=0$. 
A second case is when $p_i>p_1$ and $t_i/t_1<p_i/p_1$. 
Then the subtracted sum gives a positive contribution and 
time bins 
beyond the shortest are useful. The third case is when 
$t_i/t_1=p_i/p_1$, and this is interesting because it 
both fixes the time bins $t_i$ and leaves a degeneracy where 
a family of $n_i$ gives the same optimum. 
Thus we see that the observing strategy for this narrow 
focus on simply detections depends on the 
astrophysics of $\pburst(t)$. To explore the range of 
behaviors 
we consider three different forms for $\pburst(t)$ -- late  
activity, early activity, and peaked at some intermediate time. 
These are purely illustrative, but cover the major classes.  

We use a power law model that can 
tilt to one side or the other of the critical condition, 
and hence toward short or long times (the intermediate 
peaked case will clearly prefer an intermediate time bin, 
according to the critical condition, so we do not 
need to explore it further here), 
\be 
%\pburst(t)=\frac{p_0 m}{t_c}\,\frac{(t/t_c)^{m-1}}{(\tmax/t_c)^m}\,, 
\pburst(t)=\frac{p_0 m\, t^{m-1}}{\tmax^m}\,, 
\label{eq:burstpow} 
\ee  
where $\tmax$ is just a cutoff so $\pburst$ is normalized 
to unity (really $p_0$, the probability of there being 
any repeat burst to potentially observe) over $[0,\tmax]$. 
Then   
\be 
p_i=p_0\,\frac{t_i^m-t_{i-1}^m}{\tmax^m}\,, \label{eq:probpow} 
\ee 
and this can be used to evaluate the critical condition. 
The result is 
a transcendental equation 
\be 
\frac{t_i}{t_1}=\left(\frac{t_i}{t_1}\right)^m-1\,. 
\label{eq:critical2} 
\ee 

It is convenient to define the observing time duration ratios 
$r_{i}=t_i/t_1$. Then our main equations for the 
PDF model of Eq.~\eqref{eq:probpow} are 
\bea 
\frac{p_i}{p_1}&=&r^m_{i}-r^m_{i-1}\\ 
R&=&t_1\,\sum_i n_i r_{i}\\ 
\nflash&=&p_1\left[\frac{R}{t_1}+\sum_{i>1} n_i\,\left(r^m_{i}-r_{i}-r^m_{i-1}\right)\right]\,. \label{eq:nflcrit} 
\eea 
The quantity in parentheses in $\nflash$ will determine 
the critical values of $r_{i}$. 

Suppose $m=1$ (any burst delay time is equally likely). 
Then there is no value where the parenthetical 
vanishes, and it is always negative: 
we always have $t_i/t_1>p_i/p_1$ (i.e.\ $r_i>r_i-r_{i-1}$) 
and so the optimum 
is for all observations in the shortest time bin (i.e.\ at 
duration $t_1$). The same holds for $m<1$ since 
$r_i\ge1$. However 
for $m>1$, i.e.\ a preference for a longer delay time 
between bursts, the parenthetical can be positive, zero, or negative. 

The critical condition for Eq.~\eqref{eq:nflcrit} 
is a necessary but not sufficient condition for determining whether 
bin $i$ is preferred, i.e.\ $\nflash$ is increased by increasing 
$n_i$. It can be overruled by higher bins, depending on the 
resource constraint. Let us briefly 
consider a three bin case where bin 2 is preferred to bin 1 (so $n_1=0$), and see the impact of bin 3. 
We have  
\bea  
R&=&t_1\left(n_2 r_{2}+n_3 r_{3}\right)\\  
\nflash&=&p_1\left[R\frac{r_{3}^m-r_{2}^m}{r_{3}}+n_2\,\left\{r_{2}^m -1 -\frac{r_{2}}{r_{3}}\,\left(r_{3}^m-r_{2}^m\right)\right\}\right]\,. \label{eq:nfl23f} 
%\nflash&=&p_1\left[R\frac{r_{3}^m-r_{2}^m}{r_{3}}\right.\notag\\ &\qquad&\left.+n_2\,\left\{r_{2}^m -1 -\frac{r_{2}}{r_{3}}\,\left(r_{3}^m-r_{2}^m\right)\right\}\right]\,. \label{eq:nfl23f} 
\eea  
The quantity in curly braces is also a critical condition 
since if it is negative then the optimization gives 
$n_2=0$, i.e.\ the highest bin is preferred. 

Thus, the key criticality quantities are 
\bea 
r^\star_{i}&\quad& {\rm solves\ }\quad r^m_{i}-r_{i}-r^m_{i-1}=0 \label{eq:crit1}\\ 
r^{\star\star}_{i}&\quad& {\rm solves\ }\quad r^m_{i} -r_{i}\,\frac{r^m_{i-1}-r^m_{i-2}}{r_{i-1}}-r^m_{i-1}=0\,,  \label{eq:critf}
\eea 
and we summarize the results in Table~\ref{tab:critical}.

%%%%%%%%%%%%%%%%%%%%%%%%%% 
\begin{table}
\begin{center}
\begin{tabular}{l l}
\hline
Case\  & \ Condition \\ 
                \hline
$r_{i}<r^\star_{i}$ & 0 \\ 
$r_{i}=r^\star_{i}$ & degeneracy \\ 
$r^\star_{i}<r_{i}\le r^{\star\star}_{i}$ \ \  & 0 \\ 
$r_{i}> r^{\star\star}_{i}$ & maximum \\ 
\hline
\end{tabular}
\end{center}
\caption{Conditions for time bin $i$ to be populated or not. 
0 denotes it is unpopulated; degeneracy means that it is populated 
together with the $i-1$ bin and the numbers in each form a 
family along a linear combination degeneracy; maximum means that 
all bins $<i$ are unpopulated. The degeneracy and maximum 
conditions can be overridden by bins $>i$ such that bin $i$ 
becomes unpopulated. The critical values $r^\star_{i}$ 
and $r^{\star\star}_{i}$ are defined in Eqs.~\eqref{eq:crit1} and 
\eqref{eq:critf}. 
}
\label{tab:critical}
\end{table}

%%%%%%%%%%%%%%%%%%%%%%% 

\end{document}